\documentstyle[twocolumn,prc,aps,epsfig]{revtex}
\input epsf
\newcommand{\be}{\begin{eqnarray}}
\newcommand{\ee}{\end{eqnarray}}

 \newcommand{\gsim}{\mathrel{\hbox{\rlap{\lower.55ex \hbox {$\sim$}}
                   \kern-.3em \raise.4ex \hbox{$>$}}}}
\newcommand{\lsim}{\mathrel{\hbox{\rlap{\lower.55ex \hbox {$\sim$}}
                   \kern-.3em \raise.4ex \hbox{$<$}}}}
\newcommand{\la}{\langle}
\newcommand{\ra}{\rangle}

\def\roughly#1{\mathrel{\raise.3ex\hbox{$#1$\kern-.75em%
\lower1ex\hbox{$\sim$}}}}
\def\lsim{\roughly<}
\def\gsim{\roughly>}

\def\la{{\Big<}}
\def\ra{{\Big>}}

\begin{document}

\twocolumn[\hsize\textwidth\columnwidth\hsize\csname @twocolumnfalse\endcsname

\title{Chiral Doubling of  Heavy-Light Hadrons: \\
      BaBar 2317 MeV/$c^2$ and CLEO 2463 MeV/$c^2$ Discoveries}

\author{Maciej A. Nowak$^{a,b}$, Mannque Rho$^{c,d}$ and Ismail Zahed$^{e}$}

\address{
$^a$M. Smoluchowski Institute of Physics, Jagiellonian University,
Reymonta 4, 30-059~Krak\'ow, Poland. \\
$^b$Gesellschaft f\"{u}r Schwerionenforschung, Planckstrasse 1,
D-64291 Darmstadt, Germany. \\
$^c$ Service de Physique Th\'{e}orique, C.E. Saclay, 91191
Gif-sur-Yvette, France.\\
$^d$ School of Physics, Korea Institute for Advanced Study, Seoul
130-722, Korea.\\
 $^e$ Department of Physics and Astronomy, State
University of New York at Stony Brook, New York 11794, USA. }

\maketitle
\begin{abstract}
\noindent We point out that the very recent discoveries of BaBar
(2317)  and CLEO II (2460) are  consistent with the  general
pattern of spontaneous breaking of chiral symmetry in hadrons
built of heavy {\em and} light quarks,  as originally suggested by
us in 1992~\cite{NRZ92}, and independently by Bardeen and Hill in
1993~\cite{BH93}. The splitting between the chiral doublers
follows from a mixing between the light constituent quark mass and
the velocity of the heavy quark, and vanishes for a zero
constituent quark mass. The strictures of spontaneous
chiral symmetry breaking constrain the axial charges in the
chiral multiplet, and yield a mass splitting between the chiral
doublers of about 345 MeV when the pion coupling to the doublers
is half its coupling to a free quark. The chiral corrections are
small. This phenomenon is generic
and extends to all heavy-light hadrons. We predict the mass splitting
for the chiral doublers of the excited mesons $(D_1,D_2)$. We suggest that
the heavy-light doubling can be used to address issues of chiral
symmetry restoration in dense and/or hot hadronic matter. In
particular, the relative splitting between $D$ and $D^*$ mesons and their
chiral partners decreases in matter, with consequences on charmonium
evolution at RHIC.

\end{abstract}
\vspace{0.1in}
]
\newpage

\section{Introduction}

On April 12th 2003, the BaBar collaboration announced a narrow
peak of mass 2.317GeV $/c^2$ that decays into
$D_s^+\pi^0$~\cite{BABAR}. On May 12th 2003, the CLEO II
collaboration confirmed the BaBar result, and also
observed a second narrow peak of mass 2.46 GeV$/c^2$ in the final
$D_s^{*+}\pi^0$ state~\cite{CLEO}. Both discoveries triggered a
flurry of theoretical activity~\cite{OTHERS,BEH}, especially in
light of the first reports and the press release announcing that
the discovery is in disagreement with theoretical predictions.

In this note, we recall that actually the presence of these light
states was predicted by theoretical arguments already in 1992 and
1993, and is in fact required from the point of view of symmetries
of the QCD interactions. The two particles observed by BaBar and
CLEO II are the first chiral partners of hadrons theoretically
anticipated built out of light and heavy quarks. As such, they
represent rather a {\em pattern} of spontaneous breakdown of
chiral symmetry than isolated events.

Strong interactions involve three light flavors  (u, d, s) and
three heavy flavors (c, b, t) with respect to the QCD infrared
scale. The light sector (l) is characterized by the spontaneous
breaking of chiral symmetry, while the heavy sector (h) exhibits
heavy-quark (Isgur-Wise) symmetry \cite{IW}. In our original
work~\cite{NRZ92} we addressed the question of the form of the
heavy-light effective action in the limit where light flavors are
massless, while the heavy flavors are infinitely massive. Our
chief observation was that a consistent implementation of the
spontaneous breaking of chiral symmetry requires in addition to
the known $(0^-,1^-)$ heavy-light D-mesons, new and unknown
heavy-light chiral partners $(0^+,1^+)$ referred to as $\tilde{\rm
D}$-mesons. In the heavy-quark limit, the D$\tilde{\rm
D}$-splitting is small and of the order of the ``constituent
quark mass." Surprisingly, the approximate pattern of
spontaneous-symmetry breaking observed in light-light systems
carries even to heavy-light systems, in contrast to established
lore based on Coulomb bound states.

\section{One-Loop Results}

\begin{figure}[h]
\centerline{\epsfxsize=70mm \epsfbox{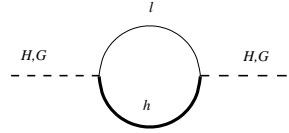}} \vskip -1.5cm
\caption{One-loop contribution  to 2-point $HH$, $GG$ functions.
Here $l$ stands for light quark and $h$ for heavy quark.}
\label{fig1}
\end{figure}

\begin{figure}[h]
\centerline{\epsfxsize=70mm \epsfbox{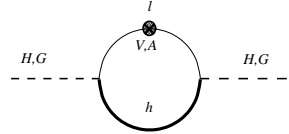}}\vskip -1.5cm
\caption{One-loop contribution to 3-point $HHV$, $HHA$, $GGV$,
$GGA$ functions. $l$ and $h$ are as in Fig.\ref{fig1}, $V$ and $A$
stand, respectively, for the external vector and axial-vector
sources.} \label{fig2}
\end{figure}

To one-loop approximation, the order $m_h^0$ contribution to the
heavy-light effective action follows from the diagrams shown in
Figs.~\ref{fig1} and~\ref{fig2} in a constituent quark
model~\footnote{We have specifically in mind the effective chiral
quark model of Manohar and Georgi~\cite{manohar-georgi}.} with
light quarks of constituent mass $\Sigma$ and heavy and
non-relativistic fields of residual mass set to zero (modulo
reparametrization invariance) and a a momentum cut-off $\Lambda$.
The result for the $(0^-,1^-)$ in the presence of vector and axial
vector currents $V,A$ is~\cite{NRZ92}

\be {{\cal L}^H}=&&-\frac{i}{2}
{\rm
  Tr} ( \bar{H} v^{\mu}\partial_{\mu}H- v^{\mu}\partial_{\mu}
  \bar{H}H)\nonumber\\
&& + {\rm Tr} V_{\mu}\bar{H}H v^{\mu} - {\bf g}_H{\rm Tr}
   A_{\mu}\gamma^{\mu}\gamma_5\bar{H}H \nonumber\\&&
-{\bf m_H}(\Sigma) \,{\rm Tr}\bar{H}\,H \label{usualcopy} \ee
where ${\bf m_H}\approx -\Sigma$ is an induced (cut-off dependent)
chiral mass reflecting the dynamical generation of mass ensuing
from spontaneously broken chiral symmetry, ${\bf g}_H$ an induced
(cut-off) dependent axial coupling and $H$ the dimension $3/2$
pseudoscalar-vector multiplet~\cite{GEORGI},

\be H=
\frac{1+v\!\!\!\!/}{2}(\gamma^{\mu}D^{*}_{\mu} +i\gamma_5 D)\,\,
\ee
with a transverse vector field, i.e. $v\cdot D^*=0$.
The Trace in (\ref{usualcopy}) is over flavor and spin.
The result is in agreement with
known results~\cite{WISE,YAN,BURDMAN} with the exception of
the chiral mass contribution missing in these works.
The origin and physical implications of the latter is
important as we now discuss.

The novel aspect of our original derivation was that consistency
with the general principles of spontaneously broken chiral
symmetry requires the introduction of {\em chiral partners} in the
form of a $(0^+,1^+)$ multiplet of pseudoscalars and
transverse vectors~\cite{NRZ92}

\be
G=
\frac{1+v\!\!\!/}{2}(\gamma^{\mu}\gamma_5\tilde{D}^{*}_{\mu} +
\tilde{D})\,\,.
\ee
To leading order in the heavy-quark mass, the one-loop effective
action for the $(0^+,1^+)$ duplicates (\ref{usualcopy}) with a key
difference in the sign of the constituent mass contribution.
Specifically~\cite{NRZ92}

 \be
{{\cal L}^G}=&&-\frac{i}{2} {\rm
  Tr}(\bar{G}v^{\mu}\partial_{\mu}G-v^{\mu}\partial_{\mu}
  \bar{G}G) \nonumber\\&&+ {\rm Tr} V_{\mu}\bar{G}G v^{\mu} - {\bf g}_G{\rm Tr}
   A_{\mu}\gamma^{\mu}\gamma_5\bar{G}G
\nonumber\\&& -{\bf m_G}(\Sigma)\,{\rm Tr}\bar{G}\,G
\label{notsousualcopy}
\ee
with the induced (cutoff dependent) chiral mass ${\bf m_G}\approx +
\Sigma$ (note the sign flip in comparison to ${\bf m_H}$).
Both chiral mass contributions are
invariant under rigid  chiral $SU(2)_L\times SU(2)_R$ and
local $SU(2)_V$ symmetry~\cite{NRZ92}.

The sign flip follows
from the $\gamma_5$ difference in the definition of the
fields $H$ and $G$, in other words the parity assignment.
Indeed, the mass contribution arising from Fig. 1 has the
generic structure (constant $H$)

\be
{\rm Tr} \left({\bf P}_2\,\frac{\rlap/{Q}+\Sigma}{Q^2-\Sigma^2}\,
H{\bf P}_3\frac{\rlap/{v}}{(v\cdot Q)}\bar{H}\right)
\label{x1}
\ee
and similarly for $H\rightarrow G$. The trace is over 4-momentum
$Q$, spin and flavor with ${\bf P}_2 = {\rm diag} (1,1,0)$ and
${\bf P}_3= {\rm diag} (0,0,1)$.
The range in $Q$ is $0<Q<\Lambda$ where $\Lambda$ is an
ultraviolet cut-off. We note that in (\ref{x1}) only the contribution

\be {\rm Tr} \left({\bf P}_2\,\frac{ \Sigma}{Q^2-\Sigma^2}\, H{\bf
P}_3\frac{\rlap/{v}}{v\cdot Q}\bar{H}\right)\,\,. \label{x2} \ee
is sensitive to the parity content of the heavy-light field since
$H\rlap/{v}=-H$ and $G\rlap/{v}=+G$. The result is a split between
the heavy-light mesons of opposite chirality. This unusual
contribution of the chiral quark mass stems from the fact that it
tags to the {\em velocity} $H\rlap/{v}\bar{H}$ of the heavy field
and is therefore sensitive to {\em parity}. It is not affected
by a shift $\Delta$ in the heavy quark mass, which amounts to the
substitution

\be
\frac{\rlap/{v}}{v\cdot Q}\rightarrow
\frac{\rlap/{v}\,v\cdot Q+\Delta}{(v\cdot Q)^2-\Delta^2}\,\,
\ee
which is seen to shift $H$ and $G$ in the same direction.
The reparametrization invariance (invariance under velocity
shifts of the heavy quark to order one) introduces mass shifts that are
parity insensitive to leading order in $1/m_h$~\cite{NZ93}.


The  $HG$-mass difference is dictated by the spontaneous breaking
of chiral symmetry, modulo the U(1) anomaly through instantons
which
 will be discussed elsewhere.
If we recall that the $H,G$ fields carry mass dimension $3/2$
through a rescaling of the complex dimension 1 fields by $\sqrt{m_h}$,
it follows from the normalizations of the kinetic and mass term
in (\ref{usualcopy}) and in (\ref{notsousualcopy}) that

\be
m_H=&&m_h+{\bf m_H}\nonumber\\
m_G=&&m_h+{\bf m_G} \label{masssplit} \ee in the chiral and
heavy-quark limit. In retrospect, this result can be arrived at
simply as follows: {\bf i)} the light quark contributes a mass
shift of order of an induced cut-off dependent constituent mass
$\Sigma$; {\bf ii)} it is repulsive in the scalars (no
$i\gamma_5$) and attractive in the pseudoscalars (with
$i\gamma_5$). In this limit, the spontaneous breakdown of chiral
symmetry enforces the mass relation~\cite{NRZ92}

\be m(\tilde{D}^*)-m(D*) = m(\tilde{D})-m(D)={\bf m_G} -{\bf m_H}
\label{split} \ee since the  dispersion relation is linear after
the heavy mass reduction. The interaction term is given by

\be {{\cal L}_{HG}}=&&+
\sqrt{\frac{{\bf g}_G}{{\bf g}_H}}\,
{\rm Tr}(\gamma_5\bar{G}H \gamma^{\mu} A_{\mu})\nonumber\\&& -
\sqrt{\frac{{\bf g}_H}{{\bf g}_G}}\,
{\rm Tr} (\gamma_5\bar{H}G\,\gamma^{\mu} A_{\mu})
\label{int}
 \ee
with no vector mixing because of parity. We note that (\ref{int})
follows from the expansion of a Dirac operator with external vector
($V$) and axial-vector ($A$) sources and is in general complex since
the Dirac operator is not self-adjoint (this point is at the origin
of flavor anomalies).

These results are expected to hold qualitatively in the presence
of non-zero current quark masses, modulo the re-summation of
standard chiral logs (chiral perturbation theory) and the U(1)
anomaly (instantons). The interaction term accounts for the strong
decay of heavy mesons via emission of Goldstone bosons
$\tilde{D}\rightarrow D\,\pi$.

In~\cite{NRZ92} we used a constituent quark model to one-loop to
estimate the pertinent parameters in (\ref{usualcopy}),
(\ref{notsousualcopy}) and (\ref{int}) which were found to be
sensitive to the cut-off procedure used in regulating the one-loop
of Fig. 1 and 2. For a general covariant cutoff, the mass
splitting for a large cutoff limit is

\be {\bf m_G}-{\bf m_H}=2\Sigma \left(1-\frac 1{4\pi^2}{\rm
ln}\frac {\Lambda}{\Sigma}\right) \,\,. \label{splitx}
 \ee
with equal and finite axial charges ${\bf g_H}\approx {\bf g_G}\approx 1/3$,
such that (\ref{int}) reduces to

\be {{\cal L}_{HG}}={\rm Tr}(\gamma_5(\bar{G}H-\bar{H}G) \gamma^{\mu} A_{\mu})
\label{intx}
 \ee
The pertinent loop integrals can be found in~\cite{NRZ92}.
All integrals were evaluated with Minkowski metric and a covariant
4-dimensional cutoff to preserve reparametrization invariance.
For a finite cutoff the results were quoted in~\cite{NRZ92},
with small effects on the splitting and somehow larger effects
on one of the axial coupling. The logarithmic sensitivity of the mass
splitting is weak but expected in field theory. The approach
advocated in \cite{NRZ92} is the Wilsonian approach with a finite
and physical cutoff $\Lambda$ to separate between the hard modes
of order $m_h$ and the soft modes. Clearly, the present results
are also sensitive to a residual  mass shift $\Delta$  in the
heavy quark mass. These effects are harder to track down in the
Wilsonian approach we have followed due to the reparametrization
invariance of the formulation. These are easier to track e.g. in
dimensional regularization scheme, however both schemes  differ
due to the  presence of strong power divergences in the loop
integrations. In general, these ambiguities  call for a first
principle calculation using lattice QCD simulations or the
instanton vacuum model~\cite{CNZ1,DIAKONOV,HEAVSHUR}.

The effects of a light
current quark mass $m_l$ can be estimated e.g. in the
aforementioned instanton vacuum model~\cite{DIAKONOV}. A simple
parametrization with good comparison to lattice data~\cite{BOWMAN}
was quoted in~\cite{MUSAKHANOV} \be \Sigma (m_l) \approx m_l
+\Sigma (\sqrt{1+(m_l/d)^2}-m_l/d) \label{massinst} \ee with
$\Sigma\approx 345\, {\rm MeV}/c^2$, $d\approx
\sqrt{0.08/2N_c}8\pi\rho/R^2\approx 198\,{\rm MeV}/c^2$ for a
standard instanton size $\rho\approx 1/3$ fm and interinstanton
distance $R\approx 1$ fm. For a strange quark mass $m_s\approx
150\, {\rm MeV}/c^2$, the second term is reduced to $\Sigma/2$,
making the combination (\ref{massinst}) weakly dependent on $m_l$
and of order $\Sigma$ all the way up to the strange quark mass.
Thus, both mass splittings are about the same for $(u,d,s)$
heavy-light mesons. The width of the non-strange heavy light
partners is however not restricted by kinematics as in the case of
$D_s$, hence these particles may be much broader and harder to detect.

\section{Goldberger-Treiman Relations}

However, in our case chiral symmetry offers further important
constraints on the spontaneous generation of mass and the ensuing
pion-$H$-$G$ interactions. This allows for model independent
relations between the cut-off dependent parameters discussed
above. Indeed, in the pure pion model discussed originally in
\cite{NRZ92} (no vector dominance) the axial-vector current
$A_\mu$ in (\ref{usualcopy},\ref{notsousualcopy},\ref{intx}) is
purely pionic and reads

\be
A_\mu=\frac i2\left(\xi\partial_\mu\xi^\dagger
-\xi^\dagger\partial_\mu\xi\right)
\label{gt1}
\ee
with $\xi=e^{i\pi/2f_\pi}$. Inserting (\ref{gt1}) into (\ref{usualcopy},
\ref{notsousualcopy},\ref{intx}) yields the pseudovector $\pi$-$HG$
 interactions

\be
{\cal L}_{\pi H} = &&\frac{{\bf g}_H}{2 f_\pi}\,
{\rm Tr}\left(\gamma^\mu\gamma^5\bar{H}H\,\partial_\mu\pi\right)
\nonumber\\
{\cal L}_{\pi G} = &&\frac{{\bf g}_G}{2 f_\pi}\,
{\rm Tr}\left(\gamma^\mu\gamma^5\bar{G}G\,\partial_\mu\pi\right)
\nonumber\\
{\cal L}_{\pi HG} = &&\frac{{\bf g}_{HG}}{2 f_\pi}\, {\rm
Tr}\left(\gamma^\mu\gamma^5(\bar{G}H-\bar{H}G)\,\partial_\mu\pi\right)
\label{gt2} \ee where for generality we introduced the axial
transition coupling ${\bf g}_{HG}$ which is 1 in (\ref{intx}).
Integrating by parts in (\ref{gt2}) and using the transversality
of the heavy-vector fields result in a single Goldberger-Treiman
relation from the last of the tree couplings in (\ref{gt2})
\footnote{The transversality of the vectors in $H$ and $G$ yields zero
pseudoscalar couplings from the first two relations in (\ref{gt2}).
This point will be clarified below.}

\be
\frac 12 (m_G-m_H)\approx
\frac 12 ({\bf m}_G-{\bf m}_H) = &&\frac{f_{\pi}\,{\bf g}_{\pi HG}}{{\bf g}_{HG}}\,\,,
\label{gt3}
\ee
with ${\bf g}_{\pi HG}$ 
the pion pseudoscalar coupling to the chiral doublet in the 
heavy and chiral limit. 
This relation was originally observed in~\cite{BH93}
up to a missing factor of $1/2$. It involves
the splitting between the even-odd partners which is less sensitive to
$\Delta$. This relations is slightly modified if vector dominance
is enforced in the heavy-light sector~\cite{NRZ92}, i.e. the rhs of
(\ref{gt3}) is divided by $(m_\rho/m_{a_1})^2$ since part of the pion
field is eaten up by the $a_1$ through a Higgs-like mechanism.
The corrections due to a finite pion mass $m_\pi$  and
a large but finite heavy quark mass $m_h$  will be
discussed below on general grounds.

For comparison, we
recall that the constituent quark mass obeys the Goldberger-Treiman
relation~\cite{manohar-georgi}

\be
\Sigma= \frac{f_\pi\,{\bf g}_{\pi qq}}{{\bf g}_A}
\label{gt4}
\ee
with ${\bf g}_A\approx 0.75$ and ${\bf g}_{\pi qq}\approx {\bf g}_{\pi NN}/3\approx 3.3$.
If we were to use ${\bf g}_{\pi HG}\approx {\bf g}_{\pi qq}/2$ and ${\bf
g}_{HG}\approx {\bf g}_A$ it follows from the last relation in
(\ref{gt3}) that the splitting in the chiral multiplet would be  one
constituent quark mass

\be
m_G-m_H\approx {\bf m}_G-{\bf m}_H\approx \Sigma\,\,, \label{gt5} \ee which
speaks for a large cutoff in (\ref{splitx}). That the pion coupling to
the chiral multiplet is 1/2 its coupling to the free light quark is
forced upon us by the BaBar and CLEO II results. This may be understood
as a sign of nontrivial screening mechanism in action in the presence of
the heavy quark, that is the pion is ``busy'' half the time with the
massive quark.

The deviations from the heavy and chiral limits to (\ref{gt3}) can be
assessed using the general framework for spontaneous breaking of chiral
symmetry developed in~\cite{hz}. Within this approach, the
one-pion reduced axial transition $\tilde{D}\rightarrow D\pi$ reads

\be &&\la D(p_2)|{\bf j}_{A\mu}^a (0) |\tilde{D}(p_1)\ra
\nonumber\\
&&=\left(\frac{(p_1-p_2)_\mu}{m_D+m_{\tilde{D}}}
\,{\bf G}_1(t) +\frac{(p_1+p_2)_\mu}{m_D+m_{\tilde{D}}}\,{\bf G}_2
(t)\right)\,D^\dagger\frac {\tau^a}2\tilde{D}
\label{op1}
\ee
where ${\bf j}^a_{A\mu}$ is the one-pion reduced axial vector current
satisfying~\cite{hz}

\be {\partial}^\mu{\bf j}_{A\mu}^a (x) = f_{\pi}\,\left(\Box
+m_\pi^2\right)\,\pi^a(x)\,\,. \label{op2} \ee The first form
factor in (\ref{op1}) is one-pion reduced, and the $D, \tilde{D}$
on the RHS are unit isospinors. For $p_1=p_2$ (at rest) the axial
charge follows from $\mu=0$ as

\be
{\bf G}_2(0)\,D^\dagger\frac{\tau^a}2\,{\tilde{D}}
\nonumber
\ee
which identifies ${\bf G}_2(0)$ with the properly normalized
axial charge in the transition matrix element.
Inserting (\ref{op2}) into (\ref{op1}) gives

\be &&\la D(p_2)|\pi^a(0) |\tilde{D}(p_1)\ra  =\frac 1{f_\pi}\frac
{1/(m_D+m_{\tilde{D}})}{m_\pi^2-t}\nonumber\\&&\times
\left(t\,{\bf G}_1(t) +(m_{\tilde{D}}^2-m_D^2)\,{\bf G}_2
(t)\right) \,D^\dagger\frac{\tau^a}2\,{\tilde{D}}\,\,. \label{op3}
\ee By definition, the $\pi$-$D\tilde{D}$ coupling is

\be \la D(p_2)|\pi^a(0) |\tilde{D}(p_1)\ra ={\bf g}_{\pi
D\tilde{D}} (t)\,\frac 1{m_\pi^2-t}
\,D^\dagger{\tau^a}\,{\tilde{D}}\,\,, \label{op4} \ee which
corresponds to

\be
{\bf g}_{\pi D\tilde{D}}\,\pi^a\,\left(\tilde{D}^\dagger\tau^a\,D +{\rm
h.c.}\right)\,\,.
\nonumber
\ee
A comparison of (\ref{op4}) to (\ref{op3}) gives at the pion
pole $t\approx m_\pi^2$

\be
f_{\pi}\,{\bf g}_{\pi D\tilde{D}} (m_\pi^2) = &&+\frac 12
(m_{\tilde{D}}-m_D)\,{\bf G}_2(m_\pi^2)\nonumber\\
&&+\frac 12 \frac {m_\pi^2}{(m_D+m_{\tilde D})} {\bf G}_1(m_\pi^2)
\label{op5} \ee which is the general form of the
Goldberger-Treiman relation for the transition amplitude
$\tilde{D}\rightarrow D\pi$. In the (double) heavy and chiral
limit it reduces to (\ref{gt3}) with the identifications ${\bf
g}_{\pi D\tilde{D}}={\bf g}_{\pi HG}$ and ${\bf G}_2(0)={\bf
g}_{HG}$. The second chiral correction in (\ref{op5}) is the
analogue of the $\pi\,N$ sigma term. In our case this amounts to a
chiral correction of order ${m_\pi^2}/{4m_h}\ll m_\pi$ to
(\ref{gt3}) which is negligible.

Similar arguments can be employed to analyze the Goldberger-Treiman
relations corresponding to the $\pi HH$ and the $\pi GG$ couplings
in (\ref{gt2}). For instance, the one-pion reduced
axial transition $D^{*}\rightarrow D\pi$ yields

\be && \la D(p_2)|{\bf j}_{A\mu}^a (0) |{D}^{*}(p_1,\epsilon)\ra
\nonumber\\&&=(\epsilon_\mu (m_D+m_{D^*}) \,{\bf
H}_1(t)\nonumber\\&& + (p_1-p_2)_\mu\,\epsilon\cdot (p_1-p_2){\bf
H}_2(t)\nonumber\\&& +
(p_1+p_2)_\mu\epsilon\cdot(p_1-p_2){\bf H}_3(t))\nonumber\\
&&\times
({m_D+m_{{D}^*}})^{-1}\,D^\dagger\frac{\tau^a}2{{D}^*}\,\,,
\label{op6} \ee where $\epsilon$ is the covariantly transverse
vector polarization of the $D^*$. Again, ${\bf H}_2$ is one-pion
reduced, and the $D$ and $D^*$ on the RHS are unit isospinors.
Using the $\pi$-$DD^*$ coupling given by

\be
\frac{{\bf g}_{\pi DD^*}}{(m_D+m_{D^*})}\,
\pi^a\,\left((\partial_\mu\,D^\dagger)\,\tau^a\,D^{*\mu}+{\rm h.c.}\right)\,\,,
\label{op6x}
\ee
which is

\be && \la D(p_2)|\pi^a (0) |{D}^{*}(p_1,\epsilon)\ra
\nonumber\\&&=\frac{{\bf g}_{\pi DD^*}}{(m_D+m_{D^*})}\,
\frac{\epsilon\cdot(p_1-p_2)}{m_\pi^2-t}
D^\dagger{\tau^a}\,{{D}^*}\,\,, \label{op6xx} \ee and a rerun of
the preceding arguments yield

\be
&&2\,f_\pi \,{\bf g}_{\pi DD^*} (m_\pi^2) =
(m_D+m_{D^*})\, {\bf H}_1(m_\pi^2) \nonumber\\&&
+ m_\pi^2\, {\bf H}_2 (m_\pi^2)
+(m_{D^*}^2-m_{D}^2){\bf H}_3(m_\pi^2)\,\,.
\label{op7}
\ee
In the heavy and chiral limit, we have

\be
f_\pi\, {\bf g}_{\pi HH} = &&m_H\,{\bf g}_{H}\nonumber\\
f_\pi\, {\bf g}_{\pi GG} = &&m_G\,{\bf g}_{G} \label{op8} \ee with
${\bf H}_1(0)={\bf g}_H$. The last relation follows from an
identical reasoning. The first relation in (\ref{op8}) was noted
by Nussinov and Wetzel~\cite{nussinov} and yields semileptonic
decay widths that are consistent with data. Equation (\ref{op7})
gives its general chiral corrections. Note that the mass of the
heavy quark $m_h$ appears explicitly in (\ref{op8}) which is the
chief reason for why these relations were not {a priori}
accessible from (\ref{gt2}) through an integration by part as
(\ref{gt2}) involves solely the soft scales \footnote{If we were
to assume an arbitrary momentum for the mass shell condition,
(\ref{op8}) could be arrived at from (\ref{gt2}) through a simple
integration by part.}. Combining (\ref{op8}) with (\ref{gt3})
leads to a relation between the various axial couplings

\be \frac{{\bf g}_{\pi GG}}{{\bf g}_G} -\frac{{\bf g}_{\pi
HH}}{{\bf g}_H} =2\,\frac{{\bf g}_{\pi HG}}{{\bf g}_{HG}}\,\,.
\label{op9} \ee The $\pi$-$GG$ and $\pi$-$HH$ couplings are fixed
by semi-leptonic decays, thereby constrain the axial charges in
(\ref{op8}-\ref{op9}). Clearly, the results (\ref{op1}-\ref{op9})
are properties of QCD and should be reproduced by {\it any}
attempt to explain the strong decay $\tilde{D}\rightarrow D\pi$,
i.e. the BaBar and CLEO results. The original
approaches~\cite{NRZ92,BH93} fulfill these constraints by
construction.

\section{BaBar and CLEO Results}

As a whole, the experimental results of BaBar and CLEO are overall
consistent with the chiral doubling proposal:

{\bf i)} The even-odd parity mass shifts are the same in the spin
$0$ and $1$ channels and of the order of the constituent quark
mass of $\sim 345$ MeV,

\be m(\tilde{D}_s^+(2316.8))- m(D_s^+)
&=& 348.3 MeV/c^2\,
\nonumber \\
m(\tilde{D}_s^+(2316.8))- m(D_s^+) &=& 350.4 \pm 1.2 \pm 1.0
 MeV/c^2\,
\nonumber \\
m(\tilde{D}_s^{*+}(2463))-m(D_s^{*+})&=& 351.6\pm 1.7\pm 1.0
 MeV/c^2\,\,,\nonumber
\ee
where we used our original ``tilde" notation, for the
two new particles. The first quote is from BaBar, while
the last two quotes are from CLEO II.

{\bf ii)} The decay widths of the strange even parity states are very
small owing to the lightness of $\Sigma$, shutting off the natural
kaon decay mode $\tilde{D}\rightarrow D K$, and operating chiefly
through the isospin violating mode $\tilde{D}\rightarrow D
(\eta\rightarrow \pi^0)$. This is overall consistent with our
interaction term (\ref{int}).

{\bf iii)} No photonic (vector) channels were found.

In~\cite{NZ93},
we further pointed out that chiral partners are also expected  for the
excited $(1^+,2^+)=(D_1, D_2)$ mesons in the form of a
$(1^-,2^-)$ chiral pair. Our prediction for the masses are:

\be
m(\tilde{D}_{s1})=2721\pm 10 MeV\nonumber\\
m(\tilde{D}_{s2})=2758\pm 10 MeV
\ee
where we used as an
input the observed BaBar and CLEO splitting
for the chiral multiplet $(0^+,1^+)$ and the  mass formulae
obtained in~\cite{NZ93}. Generalized Goldberger-Treiman relations
for the excited states can also be derived using the general
arguments presented above.

We expect a similar splitting for the non-strange heavy-light
mesons, in particular a splitting of about 368 MeV between the
$D_{u,d}$ and their chiral partners $\tilde{D}_{u,d}$. The chiral
doubling should be even more pronounced for bottom mesons, since
the $1/m_h$ corrections are three times smaller. For $m_s=150$
MeV, we expect the chiral partners of $B_s$ and $B_s^*$ to be 323
MeV heavier, while the chiral partners of $B$ and $B^*$ to be 345
MeV heavier. We note that any observation of chiral doubling for B
mesons would be a strong validation for our proposal. Indeed, in
the recently proposed alternative scenarios~\cite{OTHERS}
(multiquark states, hadronic molecules, modifications of quark
potential, unitarization) a repeating pattern from charm to bottom
calls for additional assumptions.

Bardeen and Hill~\cite{BH93} suggested a
``solvable toy field-theoretical model" and arrived at totally
analogous results for chiral partners of $D$ and $D^*$ mesons,
by using a similar one-loop calculation.
Their Nambu-Jona-Lasinio model after Fierz transformation  and to one-loop
approximation reduces to our effective action construction,
 hence the consistency
between our results and theirs. As far as we know \cite{NRZ92} and \cite{BH93}
were the only early predictions of the phenomenon of chiral doubling
for charmed and bottomed hadrons involving light quarks.
This idea was later  developed further in other papers~\cite{MANY}.

Soon after the BaBar announcement, several theoretical papers
appeared~\cite{OTHERS,BEH} suggesting a variety of explanations
for the newly observed state. In particular, Bardeen, Eichten and
Hill~\cite{BEH} adapted the effective chiral
action~\cite{NRZ92,BH93} to three light flavors, exploiting  a
constituent-quark version of Goldberger-Treiman relation and
fixing the unknown parameter of the effective Lagrangian to the
experimentally observed  splitting. The results of their
calculations which are in remarkable agreement with experiments
provide a good confirmation to our and their early suggestions
for a chiral doubling in the heavy-light sector of
QCD~\cite{NRZ92,BH93}.

Many issues regarding heavy-light systems in the QCD instanton
vacuum were discussed in \cite{CNZ1,CNZ2} including constituent
heavy-baryons such as $qqQ$ and $qQQ$ and exotics such as
$\bar{Q}\bar{q} qq$, $\bar{Q}q qqq$. In particular, it was
suggested that the heavy-light H-dibaryon $(Qqq\,Qqq)$ with
$Q=c,b$ is {\em bound} owing to the smallness of the three-body
force in the presence of the heavy quark (about 10\% the value of
the two-body force). It may even have a bound chiral partner.
Finally, it is worth pointing out that the successful treatment of
heavy-light baryons as solitons~\cite{heavybaryons,NRZBOOK} could
be readily extended to the chiral doubling now revealed in the
heavy-light systems.

\section{Summary}

In this note, we have pointed out that the newly discovered
charmed mesons by BaBar and CLEO are chiral partners of charmed
and bottomed hadrons that include at least one light quark, a
pattern suggested a decade ago~\cite{NRZ92,BH93}. The result is a
chiral splitting between the even and odd parity partners of about
a constituent quark mass as reported recently by BaBar and CLEO.
More chiral partners are expected. It may be a bit of surprise
that the pion coupling to the heavy-light chiral multiplet comes
out to be 1/2 its coupling to a free quark. The experimental results
are telling us that it should be so. Although we do not have a
rigorous argument to justify it, we conjecture that the
``screening" results since the pion is ``busy" half of the time
with the heavy quark in the chiral multiplet.  A consistent
treatment of the parity doubling in the heavy-light systems --
which can answer this as well as other questions -- can be
achieved in the QCD instanton vacuum which is parameter-free,
since the vacuum dynamics is totally fixed in the light-light
systems. We have shown that the chiral corrections are small.

QCD implies chiral Ward identities in the heavy-light systems in
the form of generalized Goldberger-Treiman relations. The even-odd
splitting is constrained by one of them. Any explanation of the
strong decay $\tilde{D}\rightarrow D\pi$ should abide by these
constraints, in particular (\ref{op5}). The chiral doubling
approach used in~\cite{NRZ92,BH93} fulfills these identities by
construction in the heavy and chiral limit. For a plausible axial
charge of unity for the $D\tilde{D}$-transition amplitude, the
observed small splitting of about 345 MeV by BaBar and CLEO is
uniquely explained by a small $\pi$-$D\tilde{D}$ coupling of about
half its value to a constituent light quark. This conclusion is
generic to QCD and should therefore be reached by all the recently
proposed alternative scenarios~\cite{OTHERS} if they were to be
viable. Chiral doubling is then an immediate consequence of rigid
chiral symmetry from quantum numbers.

Particularly relevant to the on-going effort to gain a deeper
understanding of strong interactions is the question: To what
extent can the newly discovered chiral partners shed light on the
changes of the QCD vacuum caused by external parameters such as
temperature and/or baryon density? This is an important issue in
light of the current and future experiments at RHIC and  LHC as
well as at SIS 300~\cite{SIS200}. It is also an interesting
possibility for lattice simulations. Since the chiral partners are
split by the dynamically generated chiral quark mass, it is likely
that through a chiral phase transition $D$ and $D^*$ should move
towards their chiral partners $\tilde{D}$ and $\tilde{D}^*$ to
reduce to a degenerate chiral multiplet. This should prove
particularly important for charmonium absorption/regeneration in
thermal models with medium effects. Also, this can serve as a
``litmus gauge" for the size of the chiral condensate in varying
temperature and/or density as manifested in the properties of
hadrons in hot/dense medium~\cite{BR}. In the case of the $D_s$
partners the restoration will not be complete due to the
substantial current mass of the strange quark. The restoration of
chiral symmetry in light-light systems has spurred many activities
in the past (for a recent phenomenological discussion
see~\cite{BROWNRHO}) and we expect this to extend now to the
heavy-light systems.

We are pleased that the BaBaR and CLEO II results are generating
so much excitement in both the experimental and theoretical high
energy/nuclear physics community, and it is gratifying that our
old ideas have come full circle, with so many new theoretical
venues and experimental possibilities.

\section{Note added}
After submitting the paper to the database, we became aware of the
new results of the Belle collaboration~\cite{BELLE}, announced at
FPCP 2003, on two new $c {\bar{u}}$ states $D_0^{*}\,\,(2308\pm
17\pm 15 \pm 20)$ MeV and $D_1^{*0}\,\,(2427 \pm 26 \pm 20 \pm
15)$ MeV, with the spin-parity assignment $(0^+,1^+)$. They are
likely to be the chiral partners of the nonstrange $D$ and $D^*$
$(0^-,1^-)$ multiplet. As expected they are much broader compared to the
$D_s$ states. The intriguing pattern observed, i.e., that these
two new states are almost as heavy as the corresponding strange
multiplet discovered by BaBaR and CLEO, is again in qualitative
agreement with our chiral doubling arguments above.

\section{Acknowledgments}
We would like to thank Henryk Palka for drawing our attention to
the very recent Belle results. This work was partially supported
by the Polish State Committee for Scientific Research (KBN) grant
2P03B 09622 (2002-2004), and by the US DOE grant DE-FG-88ER40388.

\end{document}